\documentclass[prl,aps,amsmath,twocolumn,showpacs]{revtex4}
\usepackage{graphicx}
\begin{document}
\title{Inelastic cotunneling through a long diffusive wire}
\author{M. V. Feigelman and A. S. Ioselevich}
\affiliation{L.D.Landau Institute for Theoretical Physics, Moscow
119334, Russia,\\
Moscow Institute of Physics and Technology, Moscow 141700,
Russia.}
\date{\today}

\begin{abstract}
We show that electron transport through a long multichannel wire,
connected to leads  by tunnel junctions, at low temperatures $T$
and voltages $V$ is dominated by inelastic cotunnelling. This
mechanism results in experimentally observed power-law dependence
of conductance on $T$ and $V$, in the diffusive regime where usual
Coulomb anomaly theory  leads to exponentially low conductance.
The power-law exponent $\alpha^*$ is proportional to the distance
between contacts $L$.

\end{abstract}
\pacs{73.63.-b, 73.23.Hk}

\maketitle

Electronic transport through nanowires  was intensively studied by many groups
in the past years.  In particular, conductance of multi-channel diffusive
nanowires with relatively poor contacts to metal terminals was
measured, cf. e.g.~\cite{experiment}. The Coulomb phenomena  play an important
role in  transport, provided the contacts between the wire and
the leads are weak. The mechanism of the
Coulomb blockade, as well as the Coulomb anomaly due to tunneling
spreading of charge, are presently well understood. In the
ballistic regime (at relatively high temperature $T$ and/or bias
voltage $V$) the Coulomb effects lead to the power-law temperature
and voltage dependence of the conductance:
\begin{eqnarray}
G\equiv dI/dV\propto V^{\alpha}\;\; \mbox{(low $T$)},\qquad
G\propto T^{\alpha}\;\; \mbox{(high $T$)},
 \label{latliq}
\end{eqnarray}
characteristic for Luttinger Liquid, while at low $T$ and $V$ --
in diffusive regime -- an exponential dependence (see
\eqref{levshaig1u} below) should be observed. The puzzle is that
the power law \eqref{latliq} is found in almost all experiments,
even in those where  the conditions for the diffusive regime seem
to be fulfilled.

The existing theories (see, e.g.,
\cite{LevitovShytov},\cite{MishchenkoAndreevGlazman},
\cite{EggerGogolin01})  considered the Coulomb effects at each of
contacts separately. However, if both contacts are taken into
account simultaneously, then some analog of cotunneling becomes
possible and at low temperatures this mechanism should dominate.
The standard theory of cotunneling deals with small grains or
quantum dots, while a long wire is an extended object: internal
dynamics of charge within it may be important. In the present
letter we develop a theory for such an extended cotunneling and
show that in the diffusive regime the resulting cotunneling
conductance still obeys the law \eqref{latliq}, though with
different exponent $\alpha^*$, depending on the separation $L$
between the contacts.

\begin{figure}
\includegraphics[width=1\columnwidth]{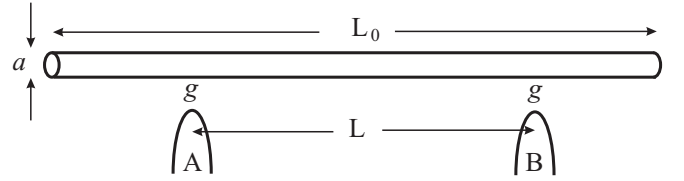}
\caption{Electrons tunnel between a wire of length $L_0$ and
diameter $a$ and two leads $A$ and $B$, placed symmetrically with
respect to the center of the wire, at distance $L$ from each
other.} \label{setup-symmetric}
\end{figure}

Consider a multichannel metallic wire (it may be a multiwall
nanotube) of length $L_0$ and diameter $a$. The wire is connected
to massive metallic leads through two weak tunnel  contacts $A$
and $B$ with identical dimensionless conductances $g\ll 1$ (see
Fig.\ref{setup-symmetric}). A voltage, applied between the
contacts is $V$.
The classic dimensionless resistance  of the piece of wire between
the leads is assumed to be not very small: $R(L)/(h/e^2)\equiv
L/\xi\gtrsim 1,$ where $\xi\sim N_{\rm ch}l$ is the localization
length,  $l$ is the mean free path,  and $N_{\rm ch}\gg 1$ is the
number of channels.

In this paper the relevant energy scales will be assumed so low,
that the motion of electrons in the wire is diffusive. On the
other hand, we will neglect the localization effects. As long as
usual conductivity in a wire is concerned, the condition  of "no
localization" reads
\begin{eqnarray}
T\gg T_{\rm Loc}\sim D\xi^{-2}\sim v_F/N_{\rm
ch}^2l.\label{cond-loc}
\end{eqnarray}
It is  not evident that inequlity~\eqref{cond-loc} is in fact necessary
when the under-the-barrier spreading process is considered; however, it is
certainly the sufficient one, and we will assume it is fulfilled below.
This requirement is consistent at $N_{\rm ch}\gg 1$  with the diffusive dynamics
of charge spreading.

If the temperature $T$ is not very low, the diffusive transport
between the two leads proceeds in a "single-particle mode": At
first one electron (one hole) tunnels into the wire from one of
the leads and is accommodated in the wire, then one hole (one
electron) tunnels from another lead.  Because of the
(thermoactivated) tunneling character of the accommodation
process, the corresponding
 conductance $G_{AB}^{(1)}$
 is exponentially suppressed
\begin{eqnarray}
G_{AB}^{(1)}\sim g\exp\{-S_1(T,V)\}.
 \label{levi1}
\end{eqnarray}
At temperatures $T\gg T_c^{(1)}\equiv\frac{E_C^2(L_0)}{E_C(\xi)}$
the accommodation proceeds according to the  semiclassic scenario
\cite{LevitovShytov} (the Coulomb zero-bias-anomaly regime, see
also \cite{AA,KamenevAndreev,MishchenkoAndreevGlazman}), and the
accommodation action
\begin{eqnarray}
S_1(T,V)\approx
\left\{\begin{aligned}0.76\sqrt{\frac{E_C(\xi)}{T}},\quad &
\mbox{for $eV\ll\sqrt{E_C(\xi)T}$,}\\
E_C(\xi)/(eV),\quad & \mbox{for $eV\gg \sqrt{E_C(\xi)T}$} .
\end{aligned}\right.
\label{levshaig1u}
\end{eqnarray}
 Here $ E_C(x)=e^2\ln (x/a)/\epsilon x$ is the
charging energy  of a piece of wire of length $x\gg a$.

At $T\ll T_c^{(1)}$ the single-particle accomodation proceeds
according the "orthodox" Coulomb blockade scenario (see, e.g.,
\cite{Devoret}):
\begin{eqnarray}
 S_1(T,V)\approx E_C(L_0)/T.
\label{levshc24p}
\end{eqnarray}

The abovementioned independent single particle  processes should
be less effective than some correlated cotunneling process, in
which any charged states of the wire would only enter as virtual
intermediate states. The theory of such processes is well
developed for transitions via small grains, where  the intergrain
charge transfer processes are the bottlenecks for the transport,
while the intragrain charge transfer is easy (see
\cite{cotunneling}). In our case, however, the charge spreading
within the wire is a crucial ingredient of the process, so that
the standard perturbational description of the cotunneling is
inapplicable.

In the present Letter we propose a modification of
the approach~\cite{LevitovShytov}, which allows for description of
charge spreading effects under the two-particle cotunneling
conditions. Our main result  reads as follows:
\begin{eqnarray}
G_{AB}^{(2)}\sim g^2 \left(\frac{\max\{LT/\xi,\,
(eV)\}}{E_C(L)}\right)^{\alpha^*},\quad  \alpha^*
=\frac{R(L)}{(h/2e^2)}, \label{ooptr}
\end{eqnarray}
In the case $L\approx L_0$ the crossover from one-particle
tunneling (in the Coulomb blockade mode) to the two-particle one
takes place at
\begin{eqnarray}
T_c^{(2)}\sim \frac{e^2\xi}{\epsilon
L_0^2}\frac{\ln(L_0/a)}{\ln\left(\frac{e^2\xi}{\epsilon
L_0^2T}\right)+\frac{\xi}{L_0}\ln(1/g)} <T_c^{(1)}.
 \label{levi2saa}
\end{eqnarray}
The last inequality becomes strong for very low conductance of
contacts, $g \ll 1$; in this situation a sequence of crossovers
may be seen with the temperature decrease: from the Coulomb
anomaly mode \eqref{levshaig1u} to the Coulomb blockade mode
\eqref{levshc24p} at $T=T_c^{(1)}$, and then to the inelastic
cotunneling regime \eqref{ooptr}  at $T=T_c^{(2)}$.

In the case $L\ll L_0$ the Coulomb blockade regime is absent, and
the crossover from the Coulomb anomaly  to the inelastic
cotunneling takes place at
\begin{eqnarray}
T_c^{(2)}\sim \frac{e^2\xi}{\epsilon
L^2}\frac{\ln(\xi/a)}{\left[\ln\left(\frac{e^2\xi}{\epsilon
L^2T}\right)+\frac{\xi}{L}\ln(1/g)\right]^2}.
 \label{levi2sbb}
\end{eqnarray}
In the nonlinear regime the crossover between the single-particle
Coulomb anomaly  and the inelastic cotunneling takes place at
\begin{eqnarray}
eV\sim (eV)_{\rm cr}\approx\frac{e^2}{\epsilon L}\ln(\xi/a).
 \label{levi2scc}
\end{eqnarray}
Thus, at low enough temperatures, the cotunneling scenario always
dominates. On the other hand, the condition \eqref{cond-loc} of
"no localization" should also be fulfilled for applicability of
the formula \eqref{ooptr}. The necessary temperature range only
exists if
\begin{eqnarray}
1\lesssim\frac{L}{\xi}\ll \sqrt{\frac{E_C(\xi)}{T_{\rm Loc}}}\sim
\left[\frac{\ln(\xi/a)}{\epsilon}N_{\rm
ch}\right]^{1/2}.\label{levshc3g}
\end{eqnarray}
i.e. the condition $N_{\rm ch} \gg 1$ is necessary.

The method of Levitov and Shytov \cite{LevitovShytov} is based on
classic equations of motion for the electron density
$\rho(x,\tau)$ and current $j(x,\tau)$ in imaginary time
$\tau=-it$. In a case of wire one can write
\begin{eqnarray}
\frac{\partial\rho}{\partial \tau}+i\frac{\partial j}{\partial
x}={\cal J}(x,\tau), \label{levshytov1}\\
j+D\frac{\partial\rho}{\partial
x}-\tilde{\sigma}\frac{\partial}{\partial x}\int
dx'\rho(x',t)\frac{1}{\epsilon |x-x'|},\label{levshytov2}
\end{eqnarray}
where $\tilde{\sigma}=e^2\xi/2\pi\hbar$ is  effective
one-dimensional conductivity, $\epsilon$ is the effective
dielectric constant, and
 $D$ is a diffusion constant. The instanton is chosen in a form of a
symmetric bounce, so that the source ${\cal J}$ in the continuity
equation \eqref{levshytov1}
 corresponds to the
injection of one extra electron into the system at time
$t_1=-i\tau_0$ at point $x=-L/2$ with its subsequent elimination
at the same point at moment $t_2=i\tau_0$: $ {\cal
J}_1=[\delta(\tau+\tau_0)-\delta(\tau-\tau_0)]\delta(x+L/2)$.

The crucial point of our approach is that we  describe
contunnelling through a diffusive wire by the same semiclassical
equations (\ref{levshytov1},\ref{levshytov2}), but with modified
source
\begin{equation}
{\cal J}_2= [\delta(\tau+\tau_0)-\delta(\tau-\tau_0)][\delta(x+L/2)-\delta(x-L/2)]
\label{J2}
\end{equation}
which describes {\em simultaneous} tunnelling of an electron and a
hole via both contacts.

 The density $\rho(x,\tau)$ should
be real and even with respect to $\tau\to -\tau$, while the
current $j(x,\tau)$ should be purely imaginary and odd.
$\rho(x,\tau)$ and $j(x,\tau)$ are defined on the interval
$-1/2T<\tau<1/2T$ and obey periodic boundary conditions. Expanding
$\rho(x,\tau)$ and $j(x,\tau)$ in Fourier series, we get
$\rho(x,\tau)=\sum_{\omega}\rho(x,\omega)\cos(\omega\tau)$ and
$j(x,\tau)=\sum_{\omega}j(x,\omega)\sin(\omega\tau)$, where the
Matsubara frequency summation, as usual for Bose excitations, runs
over even frequencies $\omega=2\pi Tn$, with  $n=0,1,2,\ldots$.

If the wire is very long ($L_0\to\infty$), the system of equations
(\ref{levshytov1},\ref{levshytov2}) can be solved by the spatial
Fourier transformation. Then, proceeding in the full analogy with
\cite{LevitovShytov}, we obtain $G_{AB}^{(2)}\approx
g^2\exp\{-S_2(T,V)\}$, where
\begin{eqnarray}
S_2(T,V)=\tilde{S}_2(T,\tau^*_0)-2eV\tau^*_0,\label{levshqqu}\\
\left.\partial\tilde{S}_2(T,\tau_0)/\partial\tau_0\right|_{\tau_0=\tau^*_0}=2eV.\label{levshqoo}
\end{eqnarray}
\begin{eqnarray}
\tilde{S}_2(T,\tau_0) =\frac{e^2}{2}\sum_{\omega, q}
 \frac{|{\cal J}_2(\omega q)|^2U_q}{(\omega+Dq^2)(\omega+\tilde{\sigma}
q^2U_q)}, \label{levsharr}\\
{\cal J}_2(\omega,q)=-4i\sin(\omega\tau_0)\sin (qL/2) ,
\label{levsh1r}\\
 U(q)\equiv{\rm
reg}\int_{-\infty}^{\infty}\frac{e^{iqx}}{\epsilon|x|}dx=\frac{2}{\epsilon}\ln\frac{1}{qa}.\label{levshrr}
\end{eqnarray}
The semiclassical method, used above, is applicable, if
$\tilde{S}_2\gg 1$. From (\ref{levsharr},\ref{levsh1r}) it is
clear, that $\partial \tilde{S}_2/\partial\tau_0=0$ for
$\tau_0=1/4T$. Therefore we conclude that
\begin{eqnarray}
\tau^*_0(T,V\to 0)=1/4T,\label{levshahk}
\end{eqnarray}
so that in the low-voltage case the summation runs only over the
odd $n=2k+1$:
\begin{eqnarray}
S_2(T,V\to 0)\approx
\nonumber\\
\sum_{k=0}^{\infty}\int\frac{dq}{2\pi}
 \frac{16e^2T\sin^2(qL/2)U_q}{(2\pi T(2k+1)+Dq^2)(2\pi T(2k+1)+\tilde{\sigma}
q^2U_q)} \nonumber
\end{eqnarray}
 There are three different temperature ranges:  $T\gg\omega_{\max}$;  $\omega_{\min}\ll T\ll \omega_{\max}$; and
  $T\ll \omega_{\min}$, where
\begin{eqnarray}
\omega_{\max}=\tilde{\sigma}L^{-2}U(L^{-1})\sim (\xi/L)E_C(L),\label{levshc2pps8}\\
\omega_{\min}\sim DL^{-2}\sim\omega_{\max}\left[N_{\rm
ch}\frac{\ln (L/a)}{\epsilon }\right]^{-1}\ll \omega_{\max}
 ,
\label{levshc2pps}
\end{eqnarray}
we consider these three ranges separately

1.  $T\gg\omega_{\max}$. Here the sum is  dominated by $k\sim 1$,
$\omega\sim T$; it can be shown that $S_2(T,V\to 0)\approx
2S_1(T,V\to 0)$, which means that in this temperature range the
two-particle process  looses a competition with the one-particle
Coulomb anomaly one.

2. $\omega_{\min}\ll T\ll\omega_{\max}$. Here the integral over
$q$ is dominated by $q\sim L^{-1}$, while the logarithmical sum
over $k$ is dominated by an interval $0<k\ll \omega_{\max}/T$, so
that
\begin{eqnarray}
S_2(T,V\to 0)\approx\frac{2L}{\xi}
\ln\left(\frac{\omega_{\max}}{T}\right), \label{levsharru1}
\end{eqnarray}
and, with the help of \eqref{levshc2pps8}, we arrive at the final
expression \eqref{ooptr}.

3. $T\ll \omega_{\min}$. In this range presumably the elastic
cotunneling should dominate. However, since $\omega_{\min}\ll
T_{\rm Loc}$, the localization effects may be important here, and
we do not discuss this regime.

In the case of high voltage one has $\tau^*_0\ll 1/T$, so that the
infrared cutoff of the frequency summation in \eqref{levsharr} is
$1/\tau_0$ instead of $T$. As a result, we arrive at the same
three regimes, but with substitution $T\to 1/\tau_0$. In
particular, in the most interesting regime 2:
\begin{eqnarray}
\tilde{S}_2(T,\tau_0)\approx\frac{2L}{\xi}\ln\left(\omega_{\max}\tau_0\right),\quad\tau_0^*=\frac{4L(eV)}{\xi},
\label{leuu1}
\end{eqnarray}
and we arrive at the high-voltage version of formula
\eqref{ooptr}.

 In the case of
a finite wire the straightforward Fourier analysis of the problem
is impossible, and one has to be more accurate.  It is important
that, for two-particle tunneling, the total charge of the wire is
always zero, so that one does not have to take into account the
effects of the total charge which otherwise would be important in
a {\it finite} system at low temperature. This simplification
allows one to follow the lines of the solution described above,
with a substitution of the Fourier analysis by decomposition over
a set of eigenfunctions of a certain linear operator.

We present the action in a general form
\begin{eqnarray}
\tilde{S}_2=\frac{e^2}{2}\sum_{\omega}\left\{-\frac{j\cdot
j}{\tilde{\sigma}\omega}+ \rho\hat{U} \rho\right\}=\nonumber\\=
\frac{e^2}{2}\sum_{\omega}{\cal
J}_2\left\{-\frac{\tilde{\sigma}}{\omega}\hat{\cal G}^+
\hat{U}\hat{\nabla}^T \hat{\nabla} \hat{U}
\hat{\cal G}+\hat{\cal G}^+ \hat{U}\hat{\cal G}\right\}{\cal J}_2=\nonumber\\
=
\frac{e^2}{2}\sum_{\omega}\frac{4\sin^2(\omega\tau_0)}{\omega}J_2
\hat{\cal G}^+\hat{U}J_2, \label{solu3}
\end{eqnarray}

\begin{eqnarray}
\rho=\hat{\cal G}{\cal J}_2,\quad
j=\tilde{\sigma}\hat{\nabla}\hat{U} \rho=
\tilde{\sigma}\hat{\nabla}\hat{U}\hat{\cal G}{\cal J}_2, \quad
\hat{\cal
G}=\left[\omega-\tilde{\sigma}\hat{\Delta}\hat{U}\right]^{-1},\nonumber
\end{eqnarray}
\begin{eqnarray}
\hat{\Delta}\equiv \frac{d^2}{dx^2},\qquad [\hat{U} \rho](x)={\rm
reg}\int_{-L_0/2}^{L_0/2}\frac{\rho(x')dx'}{\epsilon |x-x'|},
\label{solu1a}
\end{eqnarray}
The rules for regularization of the singular integral
\eqref{solu1a} are similar to those, given in
\cite{LandauLifshits}.  Within the logarithmic accuracy one can
write $[\hat{U} \rho](x)\approx U(x)\rho(x)$, where
\begin{eqnarray}
U(x)\equiv\frac{1}{\epsilon}\left\{\ln\left[\frac{2}{a}\min\left\{\left(L/2-x\right),\;\lambda(x)\right\}\right]+
\right.\nonumber\\+\left.
\ln\left[\frac{2}{a}\min\left\{\left(L/2+x\right),\;\lambda(x)\right\}\right]
\right\}, \label{solu1fo}
\end{eqnarray}
and $\lambda(x) \sim [d\ln(\rho(x))/dx]^{-1}$ is a characteristic
scale of spatial variations  of the function $\rho(x)$. The
formula \eqref{solu1fo} is applicable, if the arguments of both
logarithms are large. At $\lambda\gtrsim L_0$ \eqref{solu1fo}
coincides with the result in \cite{LandauLifshits}.

 Let us introduce
a set of normalized right-eigenvectors $\varphi_m(x)$ and
eigenvalues $v_m$ of the (non-Hermitean!) operator
$\hat{U}\hat{\Delta},$ obeying the boundary conditions $\left.d
\varphi/dx\right|_{x=\pm L_0/2}=0,$ corresponding to a vanishing
electric field at the ends of the wire. The  equation and the
normalization condition for the eigenfunctions read
\begin{eqnarray}
\hat{\Delta}\varphi_m(x)=v_m
\hat{U}^{-1}\varphi_m(x),\label{solu1owwew}\\
\int_{-L_0/2}^{L_0/2}\varphi_m(x)\hat{U}^{-1}\varphi_{m'}(x)dx=\delta_{mm'}.\label{solu1owcc}
\end{eqnarray}
There is only one zero eigenvalue $v_0=0$, the corresponding
eigenmode $\varphi_0(x)={\rm const}$ describes the equipotential
distribution of charge in the wire. For the nonzero modes
($m=1,2,\ldots$) in the leading logarithmic approximation
\begin{eqnarray}
\varphi_m(x)=\sqrt{\frac{2U_m(0)}{L_0}}\cos\left[\frac{\pi
m}{2}+\frac{\pi mx}{L_0}\right], \\
v_m=-\pi^2m^2U_m(0)/L_0^2, \label{solu1jss}
\end{eqnarray}
where $U_m(x)$ is given by \eqref{solu1fo} with
$\lambda(x)=L_0/(m+1)$.
Then, for Hermitean operator $\hat{\cal G}^+\cdot\hat{U}$, one has
\begin{eqnarray}
[\hat{\cal
G}^+\cdot\hat{U}](x,x')=\sum_m\frac{\varphi_m(x)\varphi_m(x')}{\omega-\tilde{\sigma}v_m},\label{solu1kkk}
\end{eqnarray}
so that the action can now be rewritten in the form
\begin{eqnarray}
\tilde{S}_2=\frac{e^2}{2}\sum_{\omega,
m}\frac{4\sin^2(\omega\tau_0)(J_2\cdot \varphi_m)^2}{\omega
(\omega -\tilde{\sigma}v_m)}, \label{solu1p}
\end{eqnarray}
\begin{eqnarray}
(J_2\cdot \varphi_m)= 2\sqrt{\frac{2U_m(0)}{L_0}}\sin\frac{\pi
m}{2}\sin\frac{\pi mL}{2L_0}. \label{solu1jj}
\end{eqnarray}
 Thus,   only odd $m=2p+1$, $p=0,1,\ldots$ are
relevant, and
\begin{eqnarray}
\tilde{S}_2=\sum_{\omega}\frac{16e^2\sin^2\omega\tau_0}{L_0\omega}
\sum_{p=0}^{\infty} \frac{U_{2p+1}(0)\sin^2\frac{\pi
(p+1/2)L}{L_0}}{\omega+\frac{\pi^2(2p+1)^2\tilde{\sigma}U_{2p+1}(0)}{L_0^2}}
\label{solu1hu}
\end{eqnarray}
Since  $U_m(0)$ depends on $m$ logarithmically, one can  replace
$U_m(0)\to\overline{U}\equiv U_{\overline{m}}(0)$.  Here
$\overline{m}$ is the characteristic value of $m$ (or of $2p+1$),
corresponding to those terms in \eqref{solu1hu}, that give the
principal contribution to the sum. The value of $\overline{m}$
will be found {\it a posteriori}. As a result
\begin{eqnarray}
\tilde{S}_2=\frac{16e^2\overline{U}}{L_0}\sum_{\omega}\frac{\sin^2\omega\tau_0}{\omega}
\sum_{p=0}^{\infty} \frac{\sin^2\frac{\pi
(p+1/2)L}{L_0}}{\omega+\frac{\pi^2(2p+1)^2\tilde{\sigma}\overline{U}}{L_0^2}}.
\label{solu1hu11}
\end{eqnarray}
For $V\to 0$ we have again $\tau^*_0=1/4T$. In the range of our
interest ($\omega_{\min}\ll T\ll\omega_{\max}$) one can neglect
the term $\omega$ in the last denominator of \eqref{solu1hu11} and
write
\begin{eqnarray}
S_2\approx\frac{16e^2L_0}{\pi^3\tilde{\sigma}}\sum_{k=0}^{\omega_{\max}/T}\frac{1}{2k+1}
\sum_{p=0}^{\infty} \frac{\sin^2\frac{\pi
(p+1/2)L}{L_0}}{(2p+1)^2}=\nonumber\\=\frac{2L}{\xi}\ln\left[\frac{\xi
E_C(L)}{LT}\right]. \label{solu1hunn}
\end{eqnarray}
Analyzing the series  in \eqref{solu1hunn}, we find
 $\overline{m}=\overline{2p+1}\sim L_0/L$. Hence
\begin{eqnarray}
\overline{U}\approx \frac{2}{\epsilon}\ln\frac{L}{a},\qquad
\omega_{\max}=\frac{2\pi^2\ln\frac{L}{a}}{\epsilon
L^2\tilde{\sigma}}\sim (\xi/L)E_C(L). \label{soluuuup}
\end{eqnarray}
The high-voltage modification of \eqref{solu1hunn} is obtained
exactly in the same way, as it was done for the case of infinite
wire.

Thus, in the leading logarithmic approximation, the finite-size
effects do not modify the result \eqref{ooptr}. The reason is in
the special geometry, characteristic for the two-particle
cotunneling process: The initial dipole-like distribution of
charge shrinks, so that the charges always move towards the center
of the wire, and not in the opposite direction; therefore the
presence of the ends of the wire has no effect on the process.
This is not the case for one-particle tunneling, where the charge
tends to proliferate equally in both directions, and the
size-effect is important. These problems will be discussed
elsewhere.


For interpretation of the result \eqref{ooptr} let us think of the
relevant stretch of the wire between the two contacts as a
sequence of $N=L/\xi$ grains -- small pieces of length $\xi$,
each, connected by conductances $g_{\rm eff}\sim 1$. Strictly
speaking, the perturbation theory is only applicable, if  $g_{\rm
eff}\ll 1$; however, for $g_{\rm eff}\sim 1$ it should still give
qualitatively correct estimates. Then, using the results of
\cite{FeigelmanIoselevich}, we estimate the effective conductance
of this string of grains as $G_{\rm eff}\sim g^2\left(g_{\rm
eff}N^2T^2/E_C(\xi)^2\right)^{N}\sim g^2(LT/\xi
E_C(\xi))^{2L/\xi}$, which roughly agrees with \eqref{ooptr}.

In conclusion, we have shown that low-temperature {\it diffusive}
transport through a   wire, connected to the leads by {\it two}
tunnel junctions, is realized as an inelastic cotunneling process,
the effective conductance $G_{AB}^{(2)}$ obeys a power law
 with the $L$-dependent index. Our result
\eqref{ooptr} seems to be in agreement with the data\cite{experiment}.
In contrast with the ballistic case,
where~\cite{MishchenkoAndreevGlazman,EggerGogolin01}
$\alpha\ll 1$ for $N_{\rm ch}\gg 1$, our $\alpha^*$ can be quite
large. The expression \eqref{ooptr} for $\alpha^*$ is apparently
similar to the formula $\alpha=Z(0)/(h/2e^2)$, used in the
phenomenological "environmental theory"
\cite{Bachtold,Tarkiainen}. The latter approach, however, was not
able to reproduce the correct exponential behavior \eqref{levi1}
for the single-junction setup; neither can it produce specific
 value  $R(L)$ for the effective impedance $Z(0)$.

 A financial support from the RFBR grant 06-02-16533 is
gratefully acknowledged.

\end{document}